\documentclass[journal,final]{IEEEtran}
\usepackage[T1]{fontenc}
\usepackage{verbatim}
\usepackage{amsmath}
\usepackage{amssymb}
\usepackage{graphicx}
\usepackage{tikz}
\usepackage[noadjust]{cite}
\usepackage[caption=false, font=footnotesize]{subfig}
\usepackage{cancel}

\newcommand{\partiald}[2]{\ensuremath{\frac{\partial #1}{\partial #2}}}
\definecolor{DarkBlue}{rgb}{0, 0, 0.85}
\definecolor{DarkGreen}{rgb}{0, 0.5, 0}

\newcommand\revision[1]{{#1}}
\newcommand\paperplan[2]{}
\newcommand\paperplanc[2]{}
\newcommand\togglefig[1]{#1}

\makeatletter

\providecommand{\tabularnewline}{\\}

\makeatother

\usepackage{babel}
\begin{document}
\title{Sensitivity of Single-Pulse Radar Detection to Radar State Uncertainty}
\author{Mr. Austin Costley, Dr. Randall Christensen, Dr. Robert C. Leishman,
Dr. Greg Droge\thanks{This work was supported by Air Force Research Laboratory, Wright-Patterson
Air Force Base, OH.\protect \\
A. Costley is with the Electrical and Computer Engineering Department,
Utah State University, Logan, UT 84322 USA (e-mail: adcostley@gmail.com)\protect \\
R. Christensen is with Blue Origin, Kent, WA 98032 USA (e-mail: rchristensen@blueorigin.com)\protect \\
R. Leishman is with the ANT Center, Air Force Institute of Technology,
Wright-Patterson Air Force Base, OH 45433 USA (e-mail: robert.leishman@afit.edu)\protect \\
G. Droge is with the Electrical and Computer Engineering Department,
Utah State University, Logan, UT 84322 USA (e-mail: greg.droge@usu.edu)}}
\maketitle
\begin{abstract}
Mission planners for aircraft operating under threat of detection from ground-based radar systems are often concerned with the probability of detection.
Current approaches to path planning in such environments consider the radar state (i.e. radar position and parameters) to be deterministic and known.
In practice, there is uncertainty in the radar state which induces uncertainty in the probability of detection.
This paper presents a method to incorporate the uncertainty of the radar state in a single-pulse radar detection model.
The method linearizes the radar detection model with respect to the the radar state and uses the linearized models to estimate, to the first order, the variance of the probability of detection.
The results in this paper validate the linearization using Monte Carlo analysis and illustrate the sensitivity of the probability of detection to radar state uncertainty.
\end{abstract}

\section*{Nomenclature \label{sec:nomenclature}}
\addcontentsline{toc}{section}{Nomenclature}
\begin{IEEEdescription}[\IEEEusemathlabelsep\IEEEsetlabelwidth{$p_{an}$, $p_{ae}$, $p_{ad}$}]
\item[$P_D$] Probability of detection
\item[$P_{fa}$] Probability of false alarm
\item[$\mathcal{S}$] Signal-to-noise ratio
\item[$\sigma_r$] Radar cross section ($m^2$)
\item[$R$] Range to target ($m$)
\item[$c_r$] Culmination constant of radar parameters
\item[$\boldsymbol{x_a}$] Aircraft state vector
\item[$\boldsymbol{p_a^n}$] Aircraft position vector in NED frame
\item[$\boldsymbol{\Theta_a}$] Aircraft Euler angle vector                  
\item[$p_{an}$, $p_{ae}$, $p_{ad}$] Aircraft position elements in NED frame
\item[$\phi_a$, $\theta_a$, $\psi_a$] Aircraft Euler angles (roll, pitch, yaw) 
\item[$\boldsymbol{p_r^n}$] Radar position vector in NED frame
\item[$p_{rn}$, $p_{re}$, $p_{rd}$] Radar position elements in NED frame
\item[$a$, $b$, $c$] Ellipsoid RCS parameters
\item[$\lambda$] RCS azimuth angle
\item[$\phi$] RCS elevation angle
\item[$\theta_r$] Radar detection azimuth angle 
\item[$\phi_r$] Radar detection elevation angle  
\item[$C_{aa}$] Aircraft pose covariance
\item[$\sigma_{pd}$] Standard deviation of $P_D$
\end{IEEEdescription}

\section{Introduction}
\paperplanc{Intro paragraphs / Motivation}{
    \begin{itemize}
        \item Mission planners plan paths for aircraft in environment
        \item List example missions types
        \item Reference previous work and state primary contributions
        \begin{itemize}
            \item Incorporate aircraft pose uncertainty into $P_D$
            \item Examine $P_D$ sensitivity to pose uncertainty for three analytic RCS models
        \end{itemize}
    \end{itemize}
}
Manned and unmanned aircraft are often tasked with operating under threat of detection from ground-based radar systems.
Missions executed in such environments include reconnaissance \cite{Ceccarelli_micro_uav}, radar counter-measure deployment \cite{larson_path_nodate,xiao-wei_path_2010}, and combat operations \cite{kabamba_optimal_2006}.
Planners for these missions are often concerned with the probability of being detected, which is determined by a number of factors.
The factors include the aircraft position and orientation (pose), radar position, \revision{radar parameters (e.g. power, aperture, noise factor, loss factor, etc.), and the physical characteristics of the aircraft such as the radar cross section (RCS).}

\paperplanc{Literature Review}{
    \begin{itemize}
        \item Radar detection literature review
        \item Review current radar detection path planning methods
        \item Most planning models consider simplified expressions
        \item Find examples of LUT RCS Models - discuss how are they handled/approximated
        \item Radar detection planning methods do not account for uncertainty
    \end{itemize}
}

\revision{The target detection literature provides high-fidelity single-pulse radar detection models which estimate the instantaneous probability of detection given the radar position, radar parameters, and the pose of the detected aircraft.
Marcum \cite{marcum_statistical_1960} and Swerling \cite{swerling_probability_1954} express target detection as a probability for single-pulse and fluctuating target models.
Mahafza \cite{mahafza_matlab_2003} extends the work by Marcum and Swerling to include considerations for modeling modern radar systems and common RCS models.
Most path planning algorithms use simplified radar models that do not attempt to quantify the probability of detection \cite{bortoff_path_2000,chandler_uav_2000,pachter_optimal_2001,larson_path_nodate,xiao-wei_path_2010,moore_radar_2002,mcfarland_motion_1999}.
In constrast, Kabamba \cite{kabamba_optimal_2006} quantifies the probability of tracking with a logistic function approximation, but the model abstracts away the radar-specific parameters. 
This paper will use a high-fidelity model from \cite{mahafza_matlab_2003} which quantifies the probability of detection as a function of radar parameters such as power, aperture, noise factor, and loss factor.}

\revision{A common feature of the referenced path planning and radar detection literature \cite{mcfarland_motion_1999,bortoff_path_2000,chandler_uav_2000,pachter_optimal_2001,moore_radar_2002,jun_path_2003,larson_path_nodate,kabamba_optimal_2006,xiao-wei_path_2010,xu_trajectory_2020,marcum_statistical_1960,swerling_probability_1954,mahafza_matlab_2003,bry2011rapidly,blackmore_chance-constrained_2011} is that the aircraft pose, radar position, and radar parameters are deterministic and known.
However, \cite{costley2022sensitivity} shows that moderate aircraft pose uncertainty induced significant variability in the probability of detection.
The resulting mean and variance of the probability of detection is useful for path planning \cite{bry2011rapidly,blackmore_chance-constrained_2011} and error budget analysis \cite{maybeck_stochastic_1994,farrell_aided_2008,christensen2021closedloop}.} 

\paperplanc{Contribution}{
    ~\\Extension of prevoius work includes:
    \begin{itemize}
        \item Radar position and parameter uncertainty
        \item LUT RCS model
    \end{itemize}
}

The primary contribution of this paper is an extension of the framework developed in \cite{costley2022sensitivity} to incorporate radar position and parameter uncertainty in the calculation of the probability of detection for a single-pulse radar model. 
\revision{This is accomplished by linearizing the radar detection model with respect to the radar state. 
The linearized model is used to produce a first-order approximation of the variability of the probability of detection due to uncertainties in the aircraft and radar states.
The linearization is validated using Monte Carlo analysis.
Futhermote, the sensitivity of the probability of detection to aircraft and radar state uncertainties is illustrated by evaluating the radar detection model with three levels of radar state uncertainty.}
%

The remainder of this paper is organized as follows. 
The radar detection framework presented in \cite{costley2022sensitivity} is reviewed in Section \ref{sec:RadarModel}. 
The linearization of the radar detection model with uncertainty in the radar state is derived in Section \ref{sec:LinPd}.
The results of the Monte Carlo and sensitivity analyses are provided in Section \ref{sec:Results}.

\section{Radar Detection Model \label{sec:RadarModel}}
\paperplanc{Radar Model}{
    \begin{itemize}
        \item Reference radar model used in previous work
        \item Restate equations for $P_D$, $\mathcal{S}$
        \item Provide function for $c_r$ and define variables
        \item Reference previous RCS models and introduce LUT RCS model
    \end{itemize}
}

The purpose of the radar detection model in this work is to provide an expression for the probability of detection, $P_D$, as a function of the aircraft pose, radar position, and radar parameters.
A common expression for $P_D$ uses Marcum's Q-function \cite{mahafza_matlab_2003} which is dependent on the amplitude of the radar sinusoid and contains an integral that does not have a closed form solution. 
An accepted and accurate approximation to $P_D$ provided by North \cite{north_analysis_1963} and used in \cite{mahafza_matlab_2003} is
\begin{equation}
P_{D}\approx 0.5 \times \text{erfc}\left(\sqrt{-\ln P_{fa}}-\sqrt{\mathcal{S}+0.5}\right),\label{eq:pd_approx}
\end{equation}
where $P_{fa}$ is the probability of false alarm, $\mathcal{S}$ is the signal to noise ratio, and $\text{erfc}(\cdot)$
is the complementary error function 
given by
\begin{equation}
    \text{erfc}(z) = 1 - \frac{2}{\sqrt{\pi}} \int_0^z e^{-\zeta^2} d\zeta. \label{eq:erfc}
\end{equation}

The $P_{fa}$ is considered a constant for a given radar, whereas $\mathcal{S}$ is a function of radar parameters and the pose of the target aircraft.
A general expression for the signal-to-noise ratio is given by 
\begin{eqnarray}
\mathcal{S}&=&c_r\frac{\sigma_r}{kR^4},\label{eq:SNR}
\end{eqnarray}
where $k$ is Boltzmann's constant $(1.38\times10^{-23} \; J/^{\circ}K)$, $c_r$ is the consolidated radar constant, $R$ is the range to the target, and $\sigma_r$ is the RCS.
Models for the radar constant, range to the target, and the RCS are provided in the following paragraphs.

The expression for consolidated radar constant depends on the radar type. Mahafza \cite{mahafza_matlab_2003} provides an expression for a surveillance radar as 
\begin{equation}
    c_r=\frac{P_{av}A}{16T_{0}LF}\frac{T_{sc}}{\Omega},\label{eq:c_r_surv}
\end{equation}
where $P_{av}$ is the average transmitted power, $A$ is the aperture, $T_0$ is the temperature, $T_{sc}$ is the radar scan time, and $\Omega$ is the radar search volume.

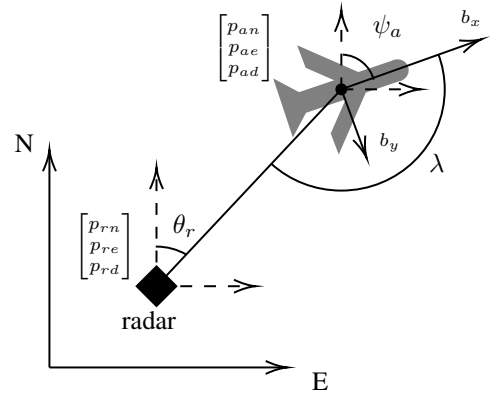
\begin{figure}
    \centering
    \tikzset{every picture/.style={line width=0.75pt}} 

\begin{tikzpicture}[x=0.75pt,y=0.75pt,yscale=-1,xscale=1]

\draw    (146,241) -- (146,133) ;
\draw [shift={(146,131)}, rotate = 90] [color={rgb, 255:red, 0; green, 0; blue, 0 }  ][line width=0.75]    (10.93,-3.29) .. controls (6.95,-1.4) and (3.31,-0.3) .. (0,0) .. controls (3.31,0.3) and (6.95,1.4) .. (10.93,3.29)   ;
\draw    (146,241) -- (264,241) ;
\draw [shift={(266,241)}, rotate = 180] [color={rgb, 255:red, 0; green, 0; blue, 0 }  ][line width=0.75]    (10.93,-3.29) .. controls (6.95,-1.4) and (3.31,-0.3) .. (0,0) .. controls (3.31,0.3) and (6.95,1.4) .. (10.93,3.29)   ;
\draw  [color={rgb, 255:red, 128; green, 128; blue, 128 }  ,draw opacity=1 ][fill={rgb, 255:red, 128; green, 128; blue, 128 }  ,fill opacity=1 ] (263.17,105.53) -- (320.08,86.24) -- (323.29,95.71) -- (266.38,115.01) -- cycle ;
\draw  [color={rgb, 255:red, 128; green, 128; blue, 128 }  ,draw opacity=1 ][fill={rgb, 255:red, 128; green, 128; blue, 128 }  ,fill opacity=1 ] (316.95,92.58) .. controls (316.06,89.96) and (317.46,87.13) .. (320.08,86.24) .. controls (322.69,85.35) and (325.53,86.75) .. (326.42,89.37) .. controls (327.31,91.98) and (325.91,94.82) .. (323.29,95.71) .. controls (320.68,96.6) and (317.84,95.2) .. (316.95,92.58) -- cycle ;
\draw  [color={rgb, 255:red, 128; green, 128; blue, 128 }  ,draw opacity=1 ][fill={rgb, 255:red, 128; green, 128; blue, 128 }  ,fill opacity=1 ] (308.1,95.19) -- (291.93,132.17) -- (288.41,122.93) -- (298.68,99.45) -- (275.63,89.38) -- (271.81,79.32) -- cycle ;
\draw  [color={rgb, 255:red, 128; green, 128; blue, 128 }  ,draw opacity=1 ][fill={rgb, 255:red, 128; green, 128; blue, 128 }  ,fill opacity=1 ] (276.17,109.95) -- (268.87,123.66) -- (259.73,96.14) -- (274.13,103.8) -- cycle ;
\draw  [dash pattern={on 4.5pt off 4.5pt}]  (293.23,100.62) -- (293.23,62.62) ;
\draw [shift={(293.23,60.62)}, rotate = 90] [color={rgb, 255:red, 0; green, 0; blue, 0 }  ][line width=0.75]    (10.93,-3.29) .. controls (6.95,-1.4) and (3.31,-0.3) .. (0,0) .. controls (3.31,0.3) and (6.95,1.4) .. (10.93,3.29)   ;
\draw  [dash pattern={on 4.5pt off 4.5pt}]  (293.23,100.62) -- (331.23,100.62) ;
\draw [shift={(333.23,100.62)}, rotate = 180] [color={rgb, 255:red, 0; green, 0; blue, 0 }  ][line width=0.75]    (10.93,-3.29) .. controls (6.95,-1.4) and (3.31,-0.3) .. (0,0) .. controls (3.31,0.3) and (6.95,1.4) .. (10.93,3.29)   ;
\draw    (293.23,100.62) -- (323.14,89.93) -- (362.12,76.01) ;
\draw [shift={(364,75.33)}, rotate = 160.34] [color={rgb, 255:red, 0; green, 0; blue, 0 }  ][line width=0.75]    (10.93,-3.29) .. controls (6.95,-1.4) and (3.31,-0.3) .. (0,0) .. controls (3.31,0.3) and (6.95,1.4) .. (10.93,3.29)   ;
\draw  [draw opacity=0] (293.99,83.13) .. controls (301.4,83.46) and (307.58,88.68) .. (309.57,95.73) -- (293.23,100.62) -- cycle ; \draw   (293.99,83.13) .. controls (301.4,83.46) and (307.58,88.68) .. (309.57,95.73) ;
\draw    (293.23,100.62) -- (200,200) ;
\draw  [draw opacity=0] (199.61,180) .. controls (199.74,180) and (199.87,180) .. (200,180) .. controls (205.39,180) and (210.46,181.42) .. (214.83,183.92) -- (200,210) -- cycle ; \draw   (199.61,180) .. controls (199.74,180) and (199.87,180) .. (200,180) .. controls (205.39,180) and (210.46,181.42) .. (214.83,183.92) ;
\draw  [draw opacity=0] (342.2,83.06) .. controls (344.25,88.54) and (345.37,94.45) .. (345.37,100.62) .. controls (345.37,128.85) and (322.02,151.73) .. (293.23,151.73) .. controls (279.68,151.73) and (267.33,146.66) .. (258.06,138.34) -- (293.23,100.62) -- cycle ; \draw   (342.2,83.06) .. controls (344.25,88.54) and (345.37,94.45) .. (345.37,100.62) .. controls (345.37,128.85) and (322.02,151.73) .. (293.23,151.73) .. controls (279.68,151.73) and (267.33,146.66) .. (258.06,138.34) ;
\draw  [color={rgb, 255:red, 0; green, 0; blue, 0 }  ,draw opacity=1 ][fill={rgb, 255:red, 0; green, 0; blue, 0 }  ,fill opacity=1 ] (200,190) -- (210,200) -- (200,210) -- (190,200) -- cycle ;
\draw  [fill={rgb, 255:red, 0; green, 0; blue, 0 }  ,fill opacity=1 ] (290.73,100.62) .. controls (290.73,99.24) and (291.85,98.12) .. (293.23,98.12) .. controls (294.61,98.12) and (295.73,99.24) .. (295.73,100.62) .. controls (295.73,102) and (294.61,103.12) .. (293.23,103.12) .. controls (291.85,103.12) and (290.73,102) .. (290.73,100.62) -- cycle ;
\draw  [dash pattern={on 4.5pt off 4.5pt}]  (200,200) -- (200,142) ;
\draw [shift={(200,140)}, rotate = 90] [color={rgb, 255:red, 0; green, 0; blue, 0 }  ][line width=0.75]    (10.93,-3.29) .. controls (6.95,-1.4) and (3.31,-0.3) .. (0,0) .. controls (3.31,0.3) and (6.95,1.4) .. (10.93,3.29)   ;
\draw  [dash pattern={on 4.5pt off 4.5pt}]  (200,200) -- (248,200) ;
\draw [shift={(250,200)}, rotate = 180] [color={rgb, 255:red, 0; green, 0; blue, 0 }  ][line width=0.75]    (10.93,-3.29) .. controls (6.95,-1.4) and (3.31,-0.3) .. (0,0) .. controls (3.31,0.3) and (6.95,1.4) .. (10.93,3.29)   ;
\draw    (293.23,100.62) -- (305.31,133.46) ;
\draw [shift={(306,135.33)}, rotate = 249.8] [color={rgb, 255:red, 0; green, 0; blue, 0 }  ][line width=0.75]    (10.93,-3.29) .. controls (6.95,-1.4) and (3.31,-0.3) .. (0,0) .. controls (3.31,0.3) and (6.95,1.4) .. (10.93,3.29)   ;

\draw (301,59.4) node [anchor=north west][inner sep=0.75pt]    {$ \begin{array}{l}
\psi _{a}\\
\end{array}$};
\draw (200,160.4) node [anchor=north west][inner sep=0.75pt]    {$ \begin{array}{l}
\theta _{r}\\
\end{array}$};
\draw (336,131.4) node [anchor=north west][inner sep=0.75pt]    {$\lambda $};
\draw (127,122) node [anchor=north west][inner sep=0.75pt]   [align=left] {N};
\draw (277,242) node [anchor=north west][inner sep=0.75pt]   [align=left] {E};
\draw (181,211) node [anchor=north west][inner sep=0.75pt]   [align=left] {radar};
\draw (227,61.4) node [anchor=north west][inner sep=0.75pt]  [font=\scriptsize]  {$\begin{bmatrix}
p_{an}\\
p_{ae}\\
p_{ad}
\end{bmatrix}$};
\draw (157,161.4) node [anchor=north west][inner sep=0.75pt]  [font=\scriptsize]  {$\begin{bmatrix}
p_{rn}\\
p_{re}\\
p_{rd}
\end{bmatrix}$};
\draw (352,59.4) node [anchor=north west][inner sep=0.75pt]  [font=\scriptsize]  {$b_{x}$};
\draw (311,121.4) node [anchor=north west][inner sep=0.75pt]  [font=\scriptsize]  {$b_{y}$};

\end{tikzpicture}    
\caption{Graphical representation of the quantities used in the radar detection model.\label{fig:radar_xy}}
\end{figure}

Consider the radar detection scenario illustrated in Fig. \ref{fig:radar_xy}.
The range to the target is defined as the 2-norm of the difference between the aircraft and radar positions in the North-East-Down (NED) frame given by
\begin{eqnarray}
    R & = & ||\boldsymbol{p_{r}^n} - \boldsymbol{p_{a}^n}||_{2}. \label{eq:range}
\end{eqnarray}
where
\begin{equation}
    \boldsymbol{p_r^n} = \begin{bmatrix} p_{rn} & p_{re} & p_{rd} \end{bmatrix}^\intercal
\end{equation}
and
\begin{equation}
    \boldsymbol{p_a^n} = \begin{bmatrix} p_{an} & p_{ae} & p_{ad} \end{bmatrix}^\intercal.
\end{equation}

\revision{The RCS of the target aircraft is a function of the angles that describe the vector from the aircraft to the radar.
These angles are referred to as the RCS azimuth angle $\lambda$ and elevation angle $\phi$ given by}
\begin{eqnarray}
    \lambda &=& \arctan\left(\frac{\rho_{ry}}{\rho_{rx}}\right) \label{eq:lambda}
\end{eqnarray}
and
\begin{eqnarray}
    \phi &=& \arctan\left(\frac{\rho_{rz}}{\sqrt{(\rho_{rx})^2+(\rho_{ry})^2}}\right), \label{eq:phi}
\end{eqnarray}
where the relative position of the radar in the body frame of the aircraft is given by
\begin{equation}
\boldsymbol{\rho_r^b} = \begin{bmatrix} \rho_{rx} & \rho_{ry} & \rho_{rz} \end{bmatrix}^\intercal.
\end{equation}
The body frame of the aircraft is defined with the x-axis out the nose of the aircraft, the z-axis out the bottom of the aircraft, and the y-axis out the right wing.
The vector $\boldsymbol{\rho_r^b}$ is calculated using the aircraft pose and radar position by
\begin{equation}
    \boldsymbol{\rho_r^b} = R_n^b \left(\boldsymbol{p_r^n}-\boldsymbol{p_a^n}\right), \label{eq:prb}
\end{equation}
where $R_n^b$ is the direction cosine matrix formed by the ZYX Euler angle sequence \cite{beard_randy_small_2012} given by
\begin{align}
    R_n^b = &\left[\begin{matrix}  C{\psi_a} C{\theta_a} & -C{\phi_a} S{\psi_a} + C{\psi_a} S{\phi_a} S{\theta_a}  \\
                C{\theta_a} S{\psi_a} & C{\phi_a} C{\psi_a} + S{\phi_a} S{\psi_a} S{\theta_a}  \\
                -S{\theta_a} & C{\theta_a} S{\phi_a} \end{matrix}\right.\nonumber \\
        & \qquad \qquad \qquad \qquad \left.\begin{matrix}
        S{\phi_a} S{\psi_a} + C{\phi_a}C{\psi_a}S{\theta_a} \\
        -C{\psi_a} S{\phi_a} + C{\phi_a} S{\psi_a} S{\theta_a} \\
        C{\phi_a} C{\theta_a} \end{matrix}\right] \label{eq:DCM_zyx}
\end{align}
where $\textrm{S}\cdot$ and $\textrm{C}\cdot$ are the $\sin(\cdot)$ and $\cos(\cdot)$ functions, and $\phi_a$, $\theta_a$, and $\psi_a$ are the Euler angles for the roll, pitch, and yaw of the aircraft, respectively.

As stated previously, the RCS is a function of $\lambda$ and $\phi$, but the specific function depends on the chosen RCS model.
The framework in \cite{costley2022sensitivity} provides support for three analytical RCS models -- constant, ellipsoid, and simple spikeball.
This paper uses the ellipsoid RCS model \cite{mahafza_matlab_2003,kabamba_optimal_2006}, although a similar derivation can be done for each of the other RCS models.
The ellipsoid RCS model represents a 3-dimensional surface given by
\begin{align}
\sigma_{re}=\frac{\pi\left(abc\right)^{2}}{\left(\left(a \,\textrm{S}\lambda \,\textrm{C}\phi\right)^{2}+\left(b \,\textrm{S}\lambda\, \textrm{S}\phi\right)^{2}+\left(c\, \textrm{C}\lambda\right)^{2}\right)^{2}} \label{eq:rcs_ellipsoid}
\end{align}
where $a$, $b$, and $c$, are the length of the ellipsoid axes.
\revision{Note that the ellipsoid orientation is aligned with the body frame with the $a$ axis forward, $b$ axis down, and $c$ axis out the right wing.}


\section{Linearized Probability of Detection \label{sec:LinPd}}
\paperplanc{Linearization Strategy}{
    \begin{itemize}
        \item Reference previous work linearization wrt aircraft pose
        \item First-order Taylor series expansion with multiple variables
        \item Compute variance using linearized models
    \end{itemize}}

This section describes a method to linearize the radar detection model with respect to the radar position and radar parameters and incorporate the linearized model into the framework in \cite{costley2022sensitivity} to account for radar state uncertainty.
The previous section has shown that $P_D$ is approximated by a nonlinear function of the aircraft pose, radar position, and radar constant $c_r$.
Thus, variability in these quantities induces variability in $P_D$.
Let the radar state be defined as
\begin{equation}
    \boldsymbol{x_r} = \begin{bmatrix} \boldsymbol{p_r^n} & c_r \end{bmatrix}^\intercal
\end{equation}
and the aircraft state be defined as
\begin{equation}
    \boldsymbol{x_a} = \begin{bmatrix}\boldsymbol{p_a^n} & \phi_a & \theta_a & \psi_a\end{bmatrix}^\intercal.
\end{equation}
A first order approximation for the mean and variance of $P_D$ is obtained by linearizing \eqref{eq:pd_approx} about the nominal aircraft and radar states and applying a Taylor series expansion.
\revision{The expansion results in the sum of $P_D$ evaluated at the nominal aircraft and radar states ($\boldsymbol{\bar{x}_a}$, $\boldsymbol{\bar{x}_r}$) with the perturbation of $P_D$, $\delta P_D$, as}  
\begin{equation}
    P_D(\boldsymbol{\bar{x}_a} + \delta \boldsymbol{x_a}, \boldsymbol{\bar{x}_r} + \delta \boldsymbol{x_r}) \approx P_D(\boldsymbol{\bar{x}_a}, \boldsymbol{\bar{x}_r}) + \delta P_D 
\end{equation}
\revision{where $\delta \boldsymbol{x_a}$ and $\delta \boldsymbol{x_r}$ are perturbations on the aircraft and radar states.}
The perturbation of $P_D$, $\delta P_D$, is expanded as
\begin{align}
    \delta P_{D} &= \frac{\partial P_D}{\partial \mathcal{S}}\begin{bmatrix}\frac{\partial \mathcal{S}}{\partial R}\frac{\partial R}{\partial\boldsymbol{x_{a}}}+\frac{\partial \mathcal{S}} {\partial\sigma_{r}}\frac{\partial\sigma_{r}}{\partial\boldsymbol{x_{a}}}\end{bmatrix} \Bigg\rvert_{\boldsymbol{\bar{x}_a}} \delta\boldsymbol{x_{a}} + \nonumber \\
                 & \qquad \partiald{P_D}{\mathcal{S}}\begin{bmatrix}\partiald{\mathcal{S}}{R}\partiald{R}{\boldsymbol{p_r^n}}+\partiald{\mathcal{S}} {\sigma_{r}}\partiald{\sigma_{r}}{\boldsymbol{p_r^n}} & \partiald{\mathcal{S}}{c_r} \end{bmatrix}\Bigg\rvert_{\boldsymbol{\bar{x}_r}} \delta\boldsymbol{x_{r}} \label{eq:pd_partial_long}\\
     & \approx A_{Pa}\delta\boldsymbol{x_{a}} + A_{Pr}\delta\boldsymbol{x_{r}}. \label{eq:pd_partial_short}
\end{align}
Thus, the perturbation can be approximated as a linear combination of perturbations on the aircraft and radar states. 
\paperplanc{$P_D$ Linearization}{
    \begin{itemize}
        \item Define partials used in \eqref{eq:pd_partial_long}
        \item Most can be obtained in other paper (should we re-define these here?)
        \item Partial of range wrt radar state $\partiald{R}{\boldsymbol{x_r}}$
        \item Partials of envelope polynomial wrt radar state $\partiald{\sigma_r}{\boldsymbol{x_r}}$
    \end{itemize}
}
The Jacobians $A_{Pa}$ and $A_{Pr}$ have three partial derivatives in common $\left(\partiald{P_D}{\mathcal{S}}, \partiald{\mathcal{S}}{R}, \partiald{\mathcal{S}}{\sigma_{r}}\right)$.
The common partial derivatives and the remaining partial derivatives that define $A_{Pa}$ are derived in \cite{costley2022sensitivity}.
The remaining partial derivatives $\left(\partiald{R}{\boldsymbol{p_r^n}}, \partiald{\sigma_r}{\boldsymbol{x_r}}, \partiald{\mathcal{S}}{c_r}\right)$ are derived in the following paragraphs. 

The partial derivative of the range to the aircraft as defined in \eqref{eq:range} with respect to the radar state is given by 
\begin{equation}
    \partiald{R}{\boldsymbol{p_r^n}} = \frac{-\left(\boldsymbol{p_a} - \boldsymbol{p_r}\right)^\intercal}{||\boldsymbol{p_a} - \boldsymbol{p_r}||_2}.
\end{equation}
\revision{The RCS, $\sigma_r$, is a function of the RCS azimuth $\lambda$ and elevation $\phi$ angles so the partial derivative of $\sigma_r$ with respect to the radar state is expanded to obtain} 
\begin{equation}
    \partiald{\sigma_r}{\boldsymbol{p_r^n}} = \partiald{\sigma_r}{\lambda} \partiald{\lambda}{\boldsymbol{p_r^n}} + \partiald{\sigma_r}{\phi} \partiald{\phi}{\boldsymbol{p_r^n}}. \label{eq:sig_r_partial}
\end{equation}
\revision{The partial derivatives of $\sigma_r$ with respect to $\lambda$ and $\phi$ $\left(\partiald{\sigma_r}{\lambda},\partiald{\sigma_r}{\phi}\right)$ are dependent on the chosen RCS model.} 
For the ellipsoid RCS model \cite{mahafza_matlab_2003,kabamba_optimal_2006,costley2022sensitivity}, these partial derivatives are given by
\begin{eqnarray}
    \partiald{\sigma_{re}}{\lambda} &=& \frac{-2 \pi (abc)^2\sin(2\lambda)\kappa}{D^3} \label{eq:dsig_dlam}\\
    \partiald{\sigma_{re}}{\phi} &=& \frac{-2 \pi (abc)^2\left(b^2-a^2\right)\sin(\lambda)^2\sin(2\phi)}{D^3}, \label{eq:dsig_dphi}
\end{eqnarray}
where
\begin{equation}
    \kappa = \left(a^2\cos(\phi)^2+b^2\sin(\phi)^2+c^2\right)
\end{equation}
and
\begin{equation}
    \textrm{D} = \left(a \,\sin\lambda \, \sin \phi\right)^{2}+\left(b \,\sin\lambda\, \sin\phi\right)^{2}+\left(c\, \cos\phi\right)^{2}.
\end{equation}
The partial derivatives of $\lambda$ and $\phi$ as defined in \eqref{eq:lambda} and \eqref{eq:phi} with respect to the radar position are common for all RCS models $\left(\partiald{\lambda}{\boldsymbol{p_r^n}}, \partiald{\phi}{\boldsymbol{p_r^n}}\right)$ and are expanded as
\begin{eqnarray}
    \partiald{\lambda}{\boldsymbol{p_r^n}} &=& \partiald{\lambda}{\boldsymbol{\rho_r^b}} \partiald{\boldsymbol{\rho_r^b}}{\boldsymbol{p_r^n}} \\
    \partiald{\phi}{\boldsymbol{p_r^n}} &=& \partiald{\phi}{\boldsymbol{\rho_r^b}} \partiald{\boldsymbol{\rho_r^b}}{\boldsymbol{p_r^n}},
\end{eqnarray}
where
\begin{eqnarray}
    \partiald{\lambda}{\boldsymbol{\rho_r^b}} &=& \begin{bmatrix} \frac{-\rho_{ry}}{\rho_{rx}^2 + \rho_{ry}^2} & 
    \frac{\rho_{rx}}{\rho_{rx}^2 + \rho_{ry}^2} &
    0 \end{bmatrix} \\
    \partiald{\phi}{\boldsymbol{\rho_r^b}} &=& \begin{bmatrix} \frac{- \rho_{rx} \rho_{rz}}{\alpha} & 
    \frac{- \rho_{ry} \rho_{rz}}{\alpha}  &
    \frac{\sqrt{\rho_{rx}^2 + \rho_{ry}^2}}{\rho_{rx}^2 + \rho_{ry}^2 + \rho_{rz}^2} \end{bmatrix}
\end{eqnarray}
and
\begin{equation}
    \alpha = \left(\rho_{rx}^2 + \rho_{ry}^2 + \rho_{rz}^2\right)\sqrt{\rho_{rx}^2 + \rho_{ry}^2}.
\end{equation}
The remaining expression is calculated by taking the partial derivative of \eqref{eq:prb} with respect to $\boldsymbol{p_r^n}$ given by
\begin{eqnarray}
    \partiald{\boldsymbol{\rho_r^b}}{\boldsymbol{p_r^n}} &=& R_n^b. 
\end{eqnarray}
Finally, the partial derivative of $\mathcal{S}$ as defined in \eqref{eq:SNR} with respect to the radar constant $c_r$ is given by
\begin{eqnarray}
    \partiald{\mathcal{S}}{c_r}&=&\frac{\sigma_r}{kR^4}.
\end{eqnarray}

The partial derivatives are used to obtain an expression for $A_{Pr}$ as
\begin{eqnarray}
    A_{Pr} &=& \partiald{P_D}{\mathcal{S}}\begin{bmatrix}\partiald{\mathcal{S}}{R}\partiald{R}{\boldsymbol{p_r^n}}+\partiald{\mathcal{S}} {\sigma_{r}}\partiald{\sigma_{r}}{\boldsymbol{p_r^n}} & \partiald{\mathcal{S}}{c_r} \end{bmatrix} \\
    &=& \frac{\exp\left(-\left(\sqrt{-\ln P_{fa}}-\sqrt{\mathcal{S}+0.5}\right)^2\right)}{2\sqrt{\pi}\sqrt{\mathcal{S}+0.5}} \nonumber \\
    & & \qquad \begin{bmatrix} \frac{4 c_r \sigma_r \left(\boldsymbol{p_a} - \boldsymbol{p_r}\right)^\intercal}{kR^{5}||\boldsymbol{p_a} - \boldsymbol{p_r}||_2} + \frac{c_r}{kR^{4}} \partiald{\sigma_{r}}{\boldsymbol{p_r^n}} & \frac{\sigma_r}{kR^4}\end{bmatrix}
\end{eqnarray}
where $\partiald{\sigma_{r}}{\boldsymbol{p_r^n}}$ is determined by \eqref{eq:sig_r_partial} for the ellipsoid RCS model.

The purpose of linearizing the radar detection model is to obtain an expression for the variability of $P_D$ due to uncertainties in the aircraft and radar states.
The variance of the linearized model is calculated by taking the expectation of $(\delta P_D)^2$ as
\begin{align}
    \sigma_{pd}^2 & = E\left[(\delta P_{D})^2\right]\\
            & = A_{Pa} C_{aa} A_{Pa}^\intercal + A_{Pr} C_{rr} A_{Pr}^\intercal \label{eq:pd_var}
\end{align}
where $C_{aa}$ is the covariance of the aircraft state and $C_{rr}$ is the covariance of the radar state.

\section{Results \label{sec:Results}}
The results of this paper are separated into two sections. 
First, results are presented to illustrate the validity of the linearization presented in Section \ref{sec:LinPd} using Monte Carlo analysis.
Second, the linearized models are used to illustrate the sensitivity of $P_D$ to variations in the aircraft and radar states.
The sensitivity of $P_D$ is examined using a moderate level of aircraft state uncertainty and three levels of radar state uncertainty (Low, Medium, and High).
The radar model used in this section is a surveillance radar from \cite{mahafza_matlab_2003} and the parameters used to generate the results are provided in Table \ref{tab:RadarParamsResults}.

\begin{table}
    \begin{centering}
    \caption{Parameters for used in results section \label{tab:RadarParamsResults}}
    \par\end{centering}
    \centering{}%
    {\renewcommand{\arraystretch}{1.25}
    \begin{tabular}{c|c|l}
    \normalsize{\textbf{Param}} & \normalsize{\textbf{Value}} & \normalsize{\textbf{Description}}\tabularnewline
    \hline 
    $a$ & 0.15 $\textrm{m}$ & Ellipsoid RCS forward axis length \tabularnewline
    $b$ & 0.13 $\textrm{m}$ & Ellipsoid RCS side axis length\tabularnewline
    $c$ & 0.21 $\textrm{m}$ & Ellipsoid RCS up axis length \tabularnewline
    ${\sigma_{pa}}$ & 10 $\textrm{m}$ & Aircraft position std. dev.\tabularnewline
    ${\sigma_{ang}}$ & 1 deg.& Euler angle state std. dev.\tabularnewline
    L ${\sigma_{pr}}$ & 10 $\textrm{m}$ & Low radar position std. dev. \tabularnewline
    M $\sigma_{pr}$ & 100 $\textrm{m}$ & Medium radar position std. dev. \tabularnewline
    H $\sigma_{pr}$ & 1000 $\textrm{m}$ & High radar position std. dev. \tabularnewline
    L $\sigma_{cr}$ & 1 J$\textrm{m}^2/^{\circ}$K & Low radar constant std. dev. \tabularnewline
    M $\sigma_{cr}$ & 5 J$\textrm{m}^2/^{\circ}$K & Medium radar constant std. dev. \tabularnewline
    H $\sigma_{cr}$ & 10 J$\textrm{m}^2/^{\circ}$K & High radar constant std. dev. \tabularnewline
    $\bar{\psi}_a$ & 90 deg. & Nominal aircraft course angle \tabularnewline
    $\bar{p}_{ad}$ & -3 $\textrm{km}$ & Nominal aircraft position - "down" axis \tabularnewline
    $\boldsymbol{\bar{p}_r^n}$ & $\boldsymbol{0_{3\times1}}$ $\textrm{m}$ & Nominal radar position vector (NED) \tabularnewline
    $\bar{c}_r$ & $167$ J$\textrm{m}^2/^{\circ}$K  & Nominal radar constant \tabularnewline
    $P_{fa}$ & $1.7e^{-4}$ & Probability of false alarm \tabularnewline
    \end{tabular}
    }
\end{table}

\subsection{Linearization Validation}
\paperplanc{Linearization Validation}{
    \begin{itemize}
        \item Describe Monte Carlo
        \item Monte Carlo hairline plots
        \begin{itemize}
            \item $P_D$ wrt aircraft state and radar state uncertainty - use ellipsoid model
            \item LUT RCS validation - show conservative bound
        \end{itemize}
        \item Summarize linearization results
    \end{itemize}
}

The linearization approach described in Section \ref{sec:LinPd} provides a first order approximation to the variation in $P_D$ due to uncertainty in the aircraft and radar states.
The validity of this approximation is dependent on the nominal operating point of the aircraft state $\boldsymbol{\bar{x}_a}$ and radar state $\boldsymbol{\bar{x}_r}$, the associated state covariances $C_{aa}$ and $C_{rr}$, and the RCS model.

The linearization is validated using Monte Carlo analysis for a scenario where the nominal radar position is at the origin of the NED frame ($\boldsymbol{\bar{p}_r^n}=\boldsymbol{0_{3\times1}}$) and nominal aircraft positions are obtained by rotating the aircraft about the radar at a nominal range.
\revision{The nominal radar constant value is $167$ $\textrm{m}^2/^{\circ}$K which is obtained from a radar example provided in \cite{mahafza_matlab_2003}.}
In this approach, the aircraft is considered at a series of nominal poses that are perturbed with random samples according to the aircraft state uncertainty level.
The radar states are similarly perturbed at the beginning of every Monte Carlo run. 
The nominal radar state is given by
\begin{eqnarray}
    \boldsymbol{\bar{x}_r} &=& \begin{bmatrix} 0 & 0 & 0 & \bar{c}_r \end{bmatrix}^\intercal. \label{eq:xrbar}
\end{eqnarray}
The $k^{th}$ nominal aircraft state is given by
\begin{equation}
    \boldsymbol{\bar{x}_a}[k] = \begin{bmatrix} \boldsymbol{\bar{p}_a^n} & 0 & 0 & \bar{\psi}_a \end{bmatrix}^\intercal
\end{equation}
where
\begin{equation}
    \boldsymbol{\bar{p}_a^n} = \begin{bmatrix} R\sin(\theta_r[k]) & R\cos(\theta_r[k]) & \bar{p}_{ad} \end{bmatrix}^\intercal
\end{equation}
and $\theta_r[k]$ ranges from 0--180 degrees in increments of 0.5 degrees.
\revision{The nominal range, $R = 500$ km, ensures that $P_D$ is near $0.5$ for some values of $\theta_r$ given the parameters defined in Table \ref{tab:RadarParamsResults}.}  
The nominal states for the $i^{th}$ Monte Carlo run are perturbed using
\begin{equation}
    \boldsymbol{x_{a,i}}[k] = \boldsymbol{\bar{x}_{a}}[k] + \boldsymbol{w_{r,i}}[k]
\end{equation}
and 
\begin{equation}
    \boldsymbol{x_{r,i}} = \boldsymbol{\bar{x}_{r}} + \boldsymbol{w_{a,i}}
\end{equation}
where $\boldsymbol{w_{a,i}}[k]$ and $\boldsymbol{w_{r,i}}$ are sampled as zero-mean Gaussian distributed random vectors with a covariance matrices given by
\begin{equation}
    C_{xx} = \begin{bmatrix} \sigma_{pa}\mathbf{I_{3 \times 3}}  & \mathbf{0} \\ \mathbf{0} & \sigma_{ang}\mathbf{I_{3 \times 3}}  \end{bmatrix}
\end{equation}
and 
\begin{eqnarray}
    C_{rr} &=& \begin{bmatrix} \sigma_{pr}\mathbf{I_{3 \times 3}} & 0 \\ \mathbf{0_{1 \times 3}} & \sigma_{cr} \end{bmatrix}.
\end{eqnarray}
The perturbed states are used to calculate $P_D[k]$ using \eqref{eq:pd_approx} for each $\theta_r[k]$ value.
The collection of $P_D$ values over the range of $\theta_r$ make up a single Monte Carlo run.
The full Monte Carlo analysis in this work consists of 500 runs.
In addition to $P_D$ calculated from the perturbed states, $\bar{P}_D$ is calculated using the nominal states, $\boldsymbol{\bar{x}_a}$ and $\boldsymbol{\bar{x}_r}$.

\begin{figure}
    \centering{}
    \togglefig{\includegraphics[width=1\columnwidth]{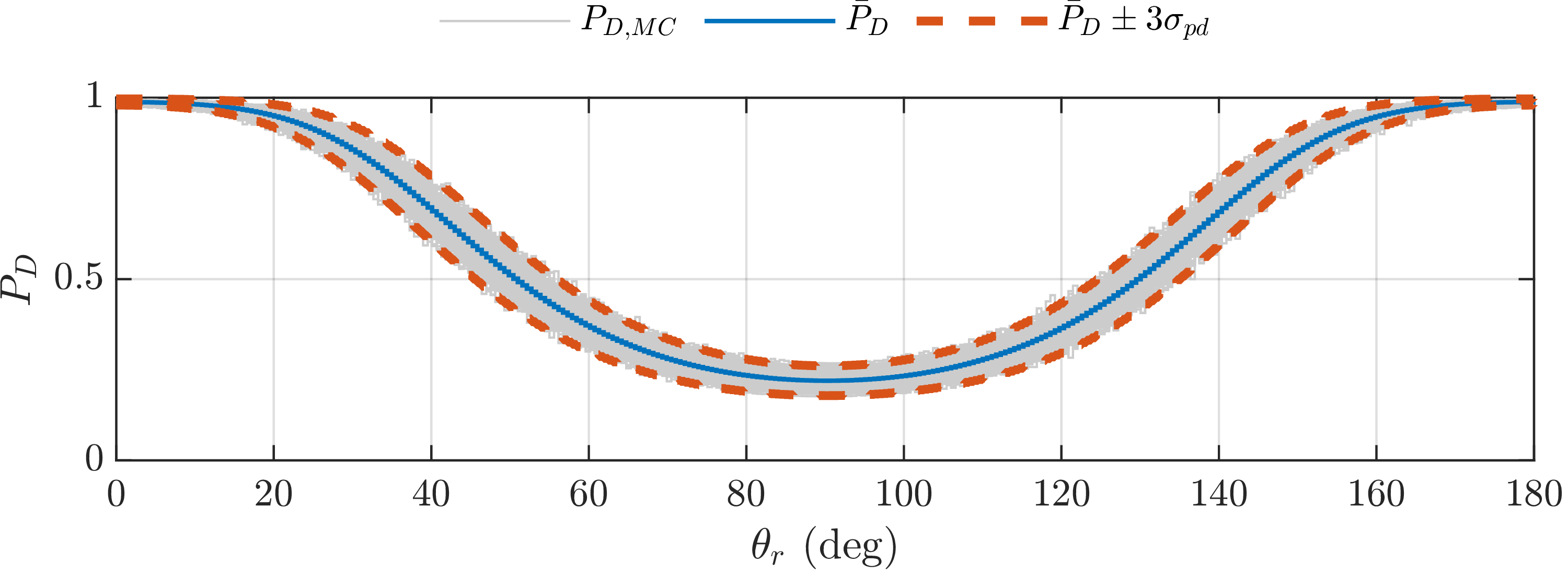}}
    \togglefig{\includegraphics[width=1\columnwidth]{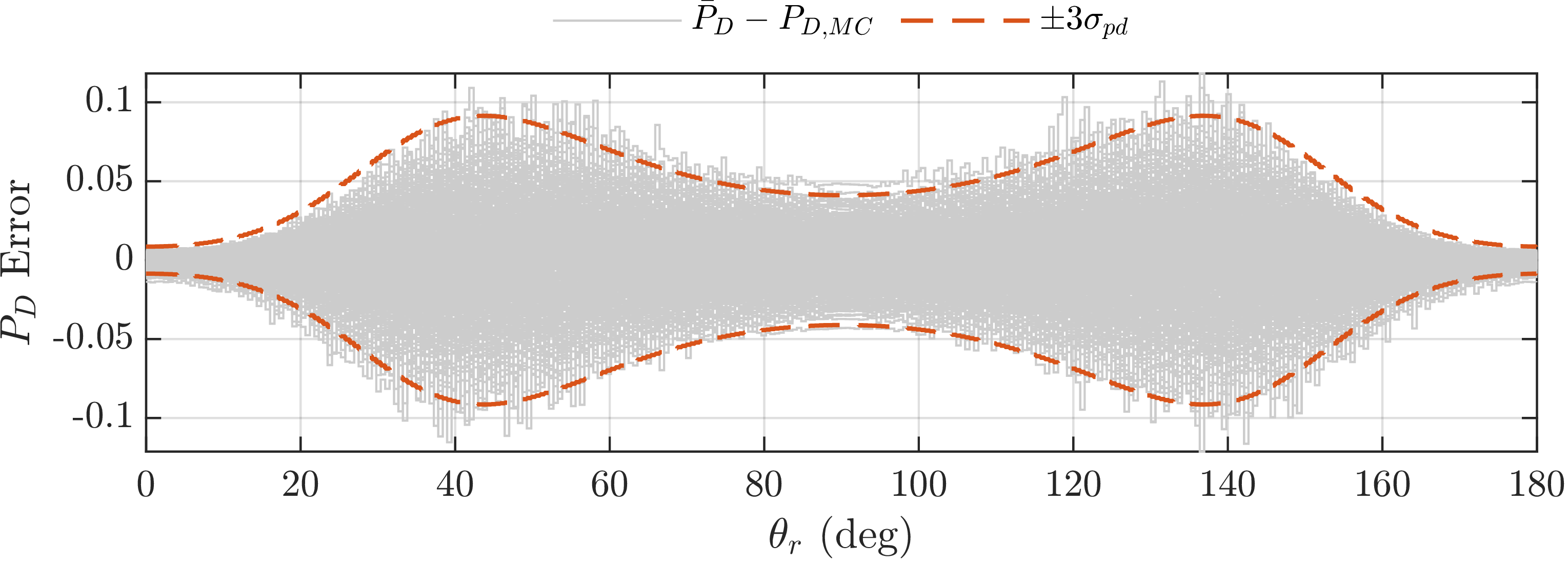}}
    \caption{Monte Carlo analysis results for $P_D$ with ``Medium'' level of aircraft and radar state uncertainty over the range $\theta_r = [0, \; 180]$  degrees with the ellipsoid RCS model. The plots show ``hair'' lines for $P_D$ (top) and $P_D$ error (bottom) for each Monte Carlo run and 3-$\sigma_{pd}$ calculated with the linearized radar model. \label{fig:mc_ellipsoid}}
\end{figure}

The first Monte Carlo analysis is performed using the ellipsoid RCS model and the Medium level of radar state uncertainty.
Fig. \ref{fig:mc_ellipsoid} shows the Monte Carlo results for this scenario.
The top plot illustrates the $P_D$ values calculated using \eqref{eq:pd_approx} for each Monte Carlo run, $P_{D,MC}$, and the nominal states, $\bar{P}_D$.
The plot also shows upper and lower 3-$\sigma_{pd}$ values as calculated using the linearized model in \eqref{eq:pd_var} which is expected to envelop $99.7\%$ of the Monte Carlo values.
The bottom plot illustrates the difference between $P_D$ calculated from the nominal states and $P_D$ calculated using \eqref{eq:pd_approx} for each Monte Carlo run. 
Observe that the 3-$\sigma_{pd}$ values of $P_D$ are nearly $0.1$ for $\theta_r \approx 45$ and $135$ degrees.
\revision{This indicates that depending on the value of $\theta_r$, the variability in $P_D$ may increase the detection risk by $0.1$.}
A key aspect to note is that the 3-$\sigma_{pd}$ values predicted by \eqref{eq:pd_var} are consistent with the ensemble statistics of the Monte Carlo runs for all values of $\theta_r$.

Two conclusions are drawn from the results in this section.
First, a moderate level of aircraft state uncertainty and Medium level of radar state uncertainty induces substantial variability in $P_D$ (nearly $\pm 0.1$ for certain detection angles).
Second, the variance of the linearized model is consistent with the extents of the ensemble statistics of the Monte Carlo runs as indicated by Fig. \ref{fig:mc_ellipsoid}.

\subsection{Sensitivity to State Uncertainty}
\paperplanc{Sensitivity Analysis}{
    \begin{itemize}
        \item Varying degrees of radar state uncertainty
        \item Error budget with aircraft position/orientation, radar position/parameter
    \end{itemize}
}

The next set of results illustrate the sensitivity of $P_D$ to uncertainty in the radar state.
The first analysis shows the 3-$\sigma$ variability of $P_D$ due to Low, Medium, and High radar state uncertainty as defined in Table \ref{tab:RadarParamsResults}.
The second analysis provides an error budget that illustrates the contribution of each of the four sources of uncertainty (i.e. aircraft position, aircraft orientation, radar position, and radar constant) to the variability of $P_D$.

\begin{figure}
    \centering{}
    \includegraphics[width=1\columnwidth]{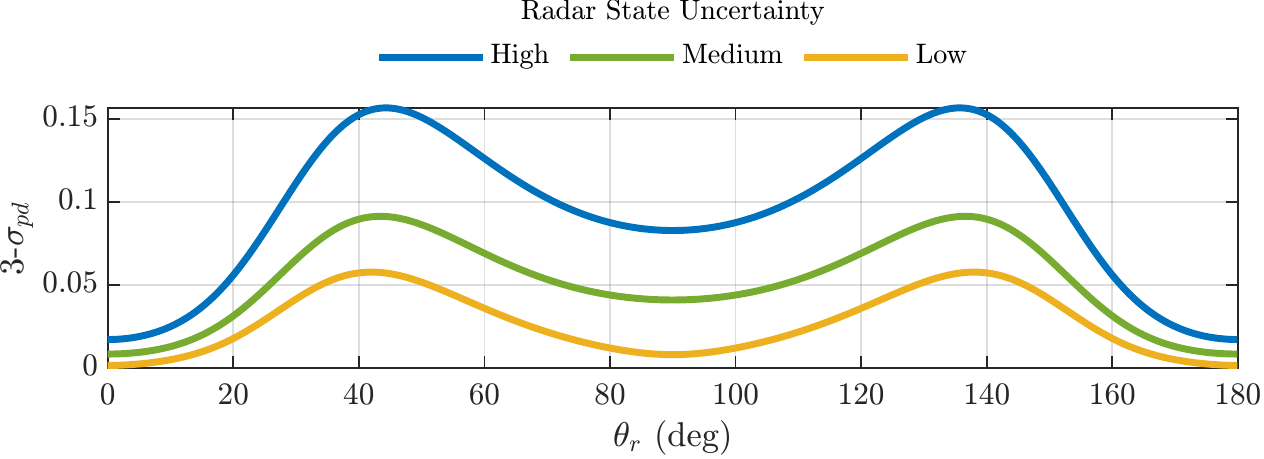}
    \caption{Sensitivity analysis of $P_D$ for three levels of radar state uncertainty over the range $\theta_r = [0, \; 180]$ degrees with the ellipsoid RCS model and $R=500$ km. Each line indicates the 3-$\sigma_{pd}$ value for each level of state uncertainty. \label{fig:sensitivity}}
\end{figure}

The first analysis is accomplished by calculating 3-$\sigma_{pd}$ using \eqref{eq:pd_var} with the nominal aircraft and radar states over a range of $\theta_r$ for the three levels of radar state uncertainty.
The nominal states were set using the same approach as the Monte Carlo analysis.
Fig. \ref{fig:sensitivity} shows the results of this analysis where, as expected, the magnitude of 3-$\sigma_{pd}$ gets larger for higher levels of radar state uncertainty.
Note that the magnitude of 3-$\sigma_{pd}$ is nearly $0.1$ for the Medium level of radar state uncertainty and over $0.15$ for the High level of radar state uncertainty for $\theta_r \approx 45$ and $135$ degrees.
This indicates that at certain detection angles, $P_D$ may be 0.1 or 0.15 higher than the nominal value due to Medium to High levels of uncertainty in the aircraft and radar states.

The second analysis provides an error budget for the Medium level of radar state uncertainty.
An error budget illustrates the contribution of each source of uncertainty (or error) in a given scenario.
This is accomplished by calculating the 3-$\sigma_{pd}$ value using \eqref{eq:pd_var} with the nominal aircraft and radar states when a single source of uncertainty is present (or activated) and the other uncertainty sources deactivated.
The method is repeated for each source of uncertainty to obtain an expression for 3-$\sigma_{pd}$ due to each source of uncertainty.

\begin{figure}
    \centering{}
    \includegraphics[width=1\columnwidth]{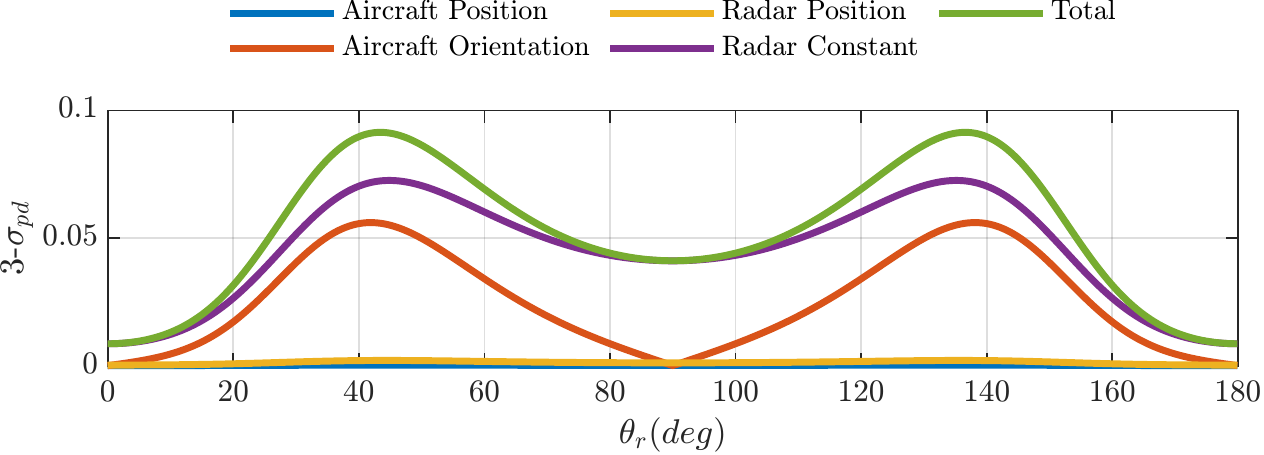}
    \caption{Error budget for variability of $P_D$ due to four sources of uncertainty for the ``Medium'' level of radar state uncertainty, ellipsoid RCS model, and $R=500$ km. Each line indicates the 3-$\sigma$ uncertainty magnitude as contributed by a given source of uncertainty. The Total line indicates the 3-$\sigma_{pd}$ magnitude for the scenario. \label{fig:error_budget}}
\end{figure}

Fig. \ref{fig:error_budget} shows the resulting error budget with four sources of uncertainty (i.e. aircraft position, aircraft orientation, radar position, and the radar constant).
The ``Total'' line shows the magnitude of 3-$\sigma_{pd}$ with all the error sources active which corresponds to the square root of the sum of the squares of the 3-$\sigma_{pd}$ values due to each contributing source.
The ``Total'' line is also the same as the ``Medium'' line from Fig. \ref{fig:sensitivity}.
Observe that the largest contributor to the overall variability in $P_D$ is the uncertainty in the radar constant when $\sigma_{cr}$ is just under 3\% of the nominal radar constant value.

Several conclusions can be drawn from the results presented in this section. 
The variability of $P_D$ is sensitive to the level of radar state uncertainty for a given scenario and the magnitude of the variability depends on the radar detection angle.
The High level of radar state uncertainty induces a 3-$\sigma_{pd}$ value of over $0.15$ for $\theta_r \approx 45$ and $135$ degrees.
These detection angles are associated with the areas of the ellipsoid RCS model with the largest variability.
Uncertainty in the aircraft and radar states has the greatest influence at these detection angles.
The contribution of individual error sources to the overall variability in $P_D$ is illustrated in the error budget in Fig. \ref{fig:error_budget}.
The graph indicates that the variability in $P_D$ is primarily driven by the uncertainty in the radar constant and the orientation of the aircraft.
If the mission planner desires to reduce the variability in $P_D$, attention should first be given to the possibility of reducing uncertainty in the radar constant and improving the estimate of the aircraft orientation.

\section{Conclusion \label{sec:Conclusion}}
\paperplanc{Summarize motivation}{}

Mission planning for aircraft operating in environments with ground-based radar systems must account for the probability of detection.
Several factors influence the probability of detection including aircraft pose, radar position, and radar performance characteristics.
Current methods for path planning in radar detection environments consider these factors to be deterministic and known.
In practice, these factors are estimated using gathered intelligence and have some uncertainty.
The methods presented in this work extend \cite{costley2022sensitivity} to incorporate uncertainty in the radar state.

\paperplanc{Summarize approach}{}

The framework in \cite{costley2022sensitivity} provides a first-order approximation of the variance of $P_D$ due to uncertainties in the aircraft state.
This is accomplished by linearizing the radar detection equations with respect to the aircraft state.
The methods in this paper extend this framework by linearizing the radar detection equations with respect to the radar state.
The linearized equations are used to compute the variability of $P_D$ due to uncertainties in the aircraft and radar states.

\paperplanc{Summarize key findings}
{
    \begin{itemize}
        \item Valid linearization using Monte Carlo
        \item LUT RCS envelopes LUT data
        \begin{itemize}
            \item Estimated $P_D$ provides conservative estimate
        \end{itemize}
        \item Signifcant variablility in $P_D$ due to variability in radar state
        \item Error budget shows radar state uncertainty is a primary driver of $P_D$ uncertainty 
    \end{itemize}
}

The linearization of $P_D$ with respect to the radar states is validated using Monte Carlo analysis for three levels of radar state uncertainty.
Two conclusions are drawn from the Monte Carlo analysis.
First, the radar state uncertainty induces significant variability in $P_D$ (nearly $0.1$ for medium radar state uncertainty).
Second, the 3-$\sigma_{pd}$ values values predicted by the linearized model are consistent with the ensemble statistics for the ellipsoid RCS model indicating valid linearization. 

The sensitivity of $P_D$ to state uncertainty is also explored in this paper.
The sensitivity analysis shows that the magnitude of the 3-$\sigma_{pd}$ values from the linearized models are over $0.15$ for High radar state uncertainty and the magnitude varies based on the detection angle. 
An error budget is generated to identify the contributions of each source of uncertainty.
The error budget indicates that the uncertainty in the radar constant is the largest contributor to the variability in $P_D$ with a radar constant standard deviation of only 3\% of the nominal value. 

The results in this paper indicate that the linearization is valid and the incorporation of the radar state uncertainty has a significant influence on the variability of $P_D$.
Failing to incorporate uncertainty in the radar state will result in detection probabilities that are higher than expected.

\bibliographystyle{IEEEtran}
\bibliography{full_bib}


\end{document}